\documentstyle[12pt]{article}
\begin{document}
\begin{titlepage}

\title { Effect of impurity on current magnification property of
mesoscopic open rings.}
\author{T. P. Pareek$^1$, P. Singha Deo$^2$ and A. M. Jayannavar$^3$.
\\%footnotemark[2] \\
Institute of Physics, Sachivalaya Marg, Bhubaneswar-751005, INDIA.}

\footnotetext[1]{e-mail:pareek@iopb.ernet.in}
\footnotetext[2]{e-mail:prosen@iopb.ernet.in}
\footnotetext[3]{e-mail:jayan@iopb.ernet.in}
\maketitle

\thispagestyle{empty}

\begin{abstract}
We have considered an open system consisting of a metallic ring
coupled to two electron reservoirs.  We have recently shown that
in the presence of a transport current, circulating currents can
flow in such a ring even in the absence of magnetic field.  This
is related to the current magnification effect in the ring. In
our present work we have studied the effect of impurity on the
current magnification.  We find that the presence of impurity
can enhance  the current magnification in the loop significantly
and thus lead to large circulating currents in certain range of
Fermi energies. This is in contrast to the known fact that
impurities can only decrease the persistent currents in a closed
ring in the presence of magnetic flux.

\end{abstract}

PACS NO :72.10.-d, 05.60.+w, 72.10.Bg, 67.57.Hi

\end{titlepage}

\eject

\newpage
\hspace {0.5in}
{\bf I. Introduction}

In the last decade physics of mesoscopic systems has emerged as
an important area of research from the point of view of basic
physics and technology[1-4]. Mesoscopic physics deals with the
structures made of metallic or semiconducting material on a
nanometer scale. The length scale associated with the system
dimensions in these systems are much smaller than the inelastic
mean free path or phase breaking length. In this regime an
electron maintains particle phase coherence across the entire
sample. In general a system with a large degrees of freedom is
called mesoscopic if the length upto which the wave function
retains phase coherence(or in general correlation length)
exceeds the size of the system.The main characteristics of
mesoscopic systems being the quantum coherence.  These systems
which are now accessible experimentally, provide an ideal
testing ground for our quantum mechanical models beyond the
atomic realm.  These systems have revealed, several interesting
new and previously unexpected quantum effects at low
temperatures[1-4], which are associated with the quantum
interference of electron waves, quantization of energy levels
and discreteness of electronic charge.

Persistent current in small metal rings threaded by magnetic
flux is prominent amongst many quantum effects in submicron
systems. B$\ddot u$ttiker et al. predicted[5] the existence of
equilibrium persistent currents in an ideal one-dimensional ring
in the presence of magnetic flux $(\phi)$.  The magnetic field
destroys the time reversal symmetry and as a consequence
persistent  current (or ring current) flow in the loop and are
periodic in magnetic flux, with a period $\phi_0$, $\phi_0$
being elementary flux quantum $(\phi_0=hc/e)$. At temperature
T=0 the amplitude of persistent current is given by $ev_{f}/L$,
where $v_f$ is the Fermi velocity and L is the circumference of
the ring. For spinless electrons persistent current is
diamagnetic or paramagnetic for N odd or even respectively.
Several recent experiments[6-8] have provided convincing
evidence that a normal metal ring threaded by a magnetic flux
exhibits in thermodynamic equilibrium a current that never
decays. After these experiments persistent current is accepted
as a parameter that characterizes equilibrium state of a small
metal ring.  However, there is a discrepancy up to two orders of
magnitude between experimental and theoretical results in the
diffusive regime. Attempts to improve upon these results has
lead to a spur in the theoretical work on persistent current in
isolated mesoscopic rings. Theoretical studies have been
extended to include multi-channel rings, and consider the
effects of disorder, spin-orbit coupling and electron-electron
interaction[9-13].  However, the experimental results obtained
in the disordered diffusive regime have not yet been explained
satisfactorily despite the intensive theoretical research.  The
problem of persistent currents has also facilitated the study of
some fundamental problems of statistical mechanics, most notably
the questions concerning the role of statistical ensemble. The
disordered average current is found to be vanishingly small for
moderate disorder when grand canonical ensemble is used, whereas
it is of finite amplitude within the framework of canonical
ensemble[1].

In contrast to the intensity of theoretical studies for isolated
systems the problem of persistent currents in open system has
received less attention[14-21]. Persistent currents occur not
only in isolated rings but also in the rings connected via leads
to electron reservoirs, namely open systems.  A simple open
system is shown in fig.(1), wherein, the metallic loop is
connected to two reservoirs characterized by chemical potentials
$\mu_1$ and $\mu_2$, respectively.  Each reservoir acts as a
source and sink for electrons and the carriers emerging from the
reservoir do not remember their earlier history. The reservoirs
absorb carrier energy and thus provides a source of energy
dissipation as well as incoherence.  All the scattering
processes in the leads including the loop are assumed to be
elastic. Inelastic processes occur only in the reservoirs, and
hence there is a complete spatial separation between elastic and
inelastic processes. Weak inelastic processes, however, do not
destroy the periodic behavior of persistent currents as a
function of magnetic flux $\phi$.  Open systems provide two
distinct possibilites: first being the equilibrium open system,
i.e., when $\mu_1=\mu_2$. The second being the case,
$\mu_1$$\neq$$\mu_2$ when a net current flows across the system
and this open system corresponds to a non-equilibrium steady
state.  It is needless to say that equilibrium closed and open
systems correspond to different statistical ensemble
descriptions namely, canonical and grand canonical ensemble
system, respectively. Due to the presence of inelastic
scattering (due to reservoirs), the amplitude of persistent
currents in equilibrium open system is smaller as compared to
the closed systems.

In our earlier studies we have pointed out that several novel
effects related to persistent currents can arise in open
systems, which have no analogue in closed or isolated systems.
In particular, we have shown that in the presence of magnetic
flux, the magnitude of persistent current is sensitive to the
direction of current flow[18], unlike the physical quantities
such as conductance. Also we have discussed the possibility  of
observing persistent current arising simultaneously due to two
non-classical effects, namely, Aharonov-Bohm effect and quantum
tunneling[17]. In our recent work we have shown that circulating
currents can arise in the presence of a transport current even
in the absence of magnetic field[19,20]. This is purely a
quantum effect and is related to the property of current
magnification in the loop. This simple quantum effect can be
explained as follows.  Consider a system of metallic loop of
circumference L coupled to two electron reservoirs,
characterized by chemical potentials $\mu_{1}$ and $\mu_{2}$
connected via ideal leads as shown in fig. (1). The leads make
the contact with the loop at junction $J_1$ and $J_2$. For
simplicity, consider a case without the $\delta$ function
impurity at X in fig.(1). The lengths of the upper and lower
arms of the loop are $l_{1}+l_2$ and $l_{3}$, respectively, such
that the length L of the circumference of the loop equals
L=$l_1+l_2+l_3$. To obtain transport current we must have
$\mu_{1}\ne\mu_{2}$ (non-equilibrium situation).  The transport
current will be directed from left to right or from right to
left depending on whether $\mu_{1}>\mu_{2}$ or
$\mu_{2}>\mu_{1}$.  The current injected[15] by the reservoir
into the lead around the small energy interval dE is given by
$I_{in}$=ev(dn/dE)f(E)dE, where v=$\hbar k/m$ is the velocity of
the carriers at the energy E, dn/dE=1/(2$\pi \hbar$v) is the
density of states in the perfect wire and f(E) is the Fermi
distribution. The total current flow I in a small energy
interval dE through the system is given by the current injected
into the leads by reservoirs multiplied by the transmission
probability T. This current splits into $I_{1}$ and $I_{2}$ in
the upper and lower arms such that I=$I_{1}+I_{2}$ (current
conservation). As the upper and lower arm lengths are unequal,
these two currents are different in magnitude.  When one
calculates quantum mechanically currents ($I_{1},I_{2}$) in the
two arms there exists two distinct possibilities. The first
possibility being for a certain range of incident Fermi wave
vectors the current in the two arms $I_{1}$ and $I_{2}$ are
individually less than the total current I, such that
I=$I_{1}+I_{2}$. In such a situation both currents in the two
arms flow in the direction of applied field. However, in certain
energy intervals, it turns out that the current in one arm is
larger than the total current I (magnification property). This
implies that to conserve the total current at the junctions the
current in the other arm must be negative, i.e., the current
should flow against the applied external field induced by the
difference in the chemical potentials. This is purely a quantum
effect.  In such a situation one can interpret the negative
current flowing in one arm, continues to flow as a circulating
current in the loop[19,20].  Thus the magnitude and direction of
circulating current is the same as that of the negative current.
Our procedure of assigning circulating current, is exactly the
same as the procedure well known in classical LCR ac network
analysis. When a parallel resonant circuit (capacitance C
connected in parallel with combination of inductance L and
resistance R) is driven by external electromotive force
(generator), circulating currents arise in the circuit at a
resonant frequency[22]. This phenomenon is well known as current
magnification.  For details we refer to ref[19,20,22]. We would
like to emphasize that in our case, for a fixed value of Fermi
energy the circulating current changes sign as we change the
direction of the current flow[19]. In equilibrium
($\mu_{1}=\mu_2$) we do not obtain any circulating current in
the absence of magnetic field. Only in the non-equilibrium
situation ($\mu_{1}\neq\mu_{2}$) , i.e, in the presence of a
steady transport current flow across the system it is possible
to observe the circulating currents.

In this paper we study the effect of impurity on the current
magnification. We have taken impurity potential to be
V(x)=$V\delta(x)$,  and the position of impurity is represented
by X in the upper arm of the loop. Our motivation to study the
impurity effect on current magnification is the following. At a
first glance we naively expect that presence of impurity leads
to increased scattering and hence suppression in current
magnification. However, we show that contrary to expectation the
current magnification can be enhanced in the presence of
impurity.

For example consider a special case of a symmetric loop (both
the arms have equal lengths) in the absence of impurity
potential i.e., $l_1+l_2=l_3$, then current magnification is not
possible. Because of the symmetry we shall have $I_1=I_2=I/2$
and currents in two arms flowing in the direction of the applied
field. Now putting an impurity in one of the arms breaks this
symmetry. Thus in general $I_1 \ne I_2$ and the current
magnification is now, possible at particular Fermi energies.  So
this simple picture tells us that impurities can enhance the
current magnification property.  Also, it should be remembered
that if we take the extreme limit of $V \rightarrow \infty$, the
current in the upper arm is zero ($I_{1}=0$), and $I_{2}=I$. In
this limit, there can be no circulating current. This simple
case suggests that impurity potential, in general, may help
enhancement of current magnification for particular Fermi energy
ranges and also can suppress amplitude of circulating currents
at some other energy ranges. In fact, we show in the following
analysis that impurity enhances current magnification
drastically for particular values of Fermi energies. The
enhanced circulating current at some energies can be as high as
10000 times the magnitude of net current flow in the system.  In
the quantum case discussed here, there is no principle
restricting the upper bound of current magnification and it can
be arbitrarily large without violating the basic law of
conservation of current at the junctions(Kirchoff's law).  In a
closed isolated loop in the presence of magnetic flux $\phi$
persistent current carried in by single particle energy level
$\epsilon_{n}$ is given by $I_n=-\partial \epsilon_n / \partial
\phi$, and the total current is I=$\Sigma I_n f(\epsilon_n)$ ,
where f($\epsilon_n$) is the Fermi function. In this case, the
presence of impurity in the otherwise ideal loop leads to
scattering, which lifts the degenaracy of states (level
repulsion) at the values $\phi=0, \pm \phi_0/2$...  etc,(at
Brillouin zone boundaries).  This in turn flattens the energy
curve as a function of $\phi$ and as a result the amplitude of
persistent current always decreases with an increase in the
impurity strength (for details see ref[9]).

{\bf II Theoretical treatment}

We now consider a case of one-dimensional metal loop of length L
coupled to two electron reservoirs as shown in fig. 1 and
current is injected into the metal loop from left
($\mu_{1}>\mu_{2}$). Our calculation is for noninteracting
system of electrons. Except at $\delta$ potential impurity (of
strength $V$) at site X the potential throughout the network is
zero (free electron network). We do not assume any particular
form for the scattering matrix for the junctions $J_1$ and
$J_2$, but scattering at the junctions follow from the first
principles using quantum mechanics.  At temperature zero the
total current flow around a small energy interval dE around E is
I=(e/$2\pi\hbar$)TdE, where T is the transmission coefficient
calculated at the energy E.  It is a straight forward exercise
to set up a scattering problem and calculate the transmission
coefficient (T) and the current densities in the upper ($I_{1}$)
and the lower ($I_{2}$) arms. We follow our earlier method of
quantum waveguide transport on networks closely to calculate
these quantities[17-20,23-25]. We have imposed the Griffiths
boundary condition (conservation of current) and single
valuedness of the wavefunctions at the junctions.  The final
analytical for the current densities expressions are given by

\begin{equation}
\frac{dI}{dE}=(e/2\pi\hbar)T,
\end{equation}

$$T=16\left\{ {V^2} {\cos{^2}[k (l_{1} - l_{2})]} -
  2 {V^2} \cos[k (l_{1} - l_{2})] \cos[k (l_{1} + l_{2})] + \right. $$
$$  {V^2} {\cos^{2}[k (l_{1} + l_{2})]} +
  4 k V \cos[k (l_{1} - l_{2})] \sin[k (l_{1} + l_{2})] -
  4 k V \cos[k (l_{1} + l_{2})] \sin[k (l_{1} + l_{2})] + $$
$$  4 {k^2} {\sin^{2}[k (l_{1} + l_{2})]} +
  4 k V \cos[k (l_{1} - l_{2})] \sin[k l_{3}] -
  4 k V \cos[k (l_{1} + l_{2})] \sin[k l_{3}] + $$
\begin{equation}
\left.  8 {k^2} \sin[k (l_{1} +  l_{2})] \sin[k l_{3}] +
  4 {k^2} \sin^{2}[k l_{3}]\right\}/\Omega,
\end{equation}

\begin{eqnarray}
\frac{dI_{1}}{dE}=(e/2\pi\hbar)16k\left\{  2 k - 2 k \cos (2 k l_{3}) +
    2 k \cos (k r) -
    2 k \cos (k L) \right. \nonumber\\
\left.  -  V \sin (k p)
    +  V \sin (k r)
    +  V \sin (k s)
    -  V \sin (k L)\right\}/\Omega,
\end{eqnarray}

\begin{eqnarray}
\frac{dI_{2}}{dE}=-(e/2\pi\hbar)16\left\{ -2 {k^2} -
     {V^2} +  {V^2} \cos[2 k l_{1}]
     - {V^2} \cos[2 k  l_{1}]\cos[2 k l_{2}]
     +  {V^2} \cos[2 k l_{2}]\right .   \nonumber\\
\left. + 2 {k^2} \cos[2 k ( l_{1} + l_{2})]
      - 2 {k^2} \cos[k r]
    +2 {k^2} \cos[k L]
- 2 k V \sin[2 k l_{1}] - 2 k V \sin[2 k l_{2}]\right .\nonumber\\
\left . + 2 k V \sin[2 k ( l_{1} + l_{2})]
     +  k V \sin[k p]
 -  k V \sin[k r]
   -  k V \sin[k s]
     +  k V \sin[k L] \right\}/\Omega ,
\end{eqnarray}

\newpage
\noindent where

 $$\Omega=\left\{64 {k^2} + 4 {V^2} \cos^{2}[k  p] +
  32 {k^2} \cos[k  r] +
  4 {k^2} \cos^{2}[k  r] +  32 {V^2} \cos[k  p] \cos[k  s] + \right.$$
$$  4 {V^2} \cos^{2}[k  s] -
  160 {k^2} \cos[k  L] -
  16 {V^2} \cos[k  p]
   \cos[k  L] - $$
$$  40 {k^2} \cos[k  r]
   \cos[k  L] -
  16 {V^2} \cos[k  s]
   \cos[k  L] +
  100 {k^2} \cos^{2}[k  L] + $$
$$ 16 {V^2} \cos^{2}[k  L] -
 16 k V \sin[k  p] -
  4 k V \cos[k  r]
   \sin[k  p] + $$
$$ 20 k V \cos[k  L]
   \sin[k  p] +
   {V^2} \sin^{2}[k  p] +
  16 k V \sin[k  r] + $$
$$  4 k V \cos[k  r]
   \sin[k  r] -
  20 k V \cos[k  L]
   \sin[k  r] -
  2 {V^2} \sin[k  p]
   \sin[k  r] + $$
$$  {V^2} \sin^{2}[k  r] +
 16 k V \sin[k  s] +
  4 k V \cos[k  r]
   \sin[k  s] - $$
$$  20 k V \cos[k  L]
   \sin[k  s] -
  2 {V^2} \sin[k  p]
   \sin[k  s] + $$
$$ 2 {V^2} \sin[k  r]
   \sin[k  s] +
   {V^2} \sin^{2}[k  s] -
  80 k V \sin[k  L] + $$
$$  32 k V \cos[k  p]
   \sin[k  L] -
  20 k V \cos[k  r]
   \sin[k  L] + $$
$$  32 k V \cos[k  s]
   \sin[k  L] +
  36 k V \cos[k  L]
   \sin[k  L] + $$
$$  10 {V^2} \sin[k  p]
   \sin[k  L] -
  10 {V^2} \sin[k  r]
   \sin[k  L] - $$
\begin{equation}
\left.  10 {V^2} \sin[k  s]
   \sin[k  L] +
  64 {k^2} \sin^{2}[k  L] +
  25 {V^2} \sin^{2}[k  L]\right\}.
\end{equation}

In the above equations k=$\sqrt{E}$, is the incident wave vector,
$p=l_{1}-l_{2}-l_{3}$,
$L=l_{1}+l_{2}+l_{3}$,
$r=l_{1}+l_{2}-l_{3}$ and
$s=l_{1}-l_{2}+l_{3}$.
We have rescaled current densities in a dimensionless form
and henceforth we denote I$(\equiv \frac{2\pi\hbar}{e}\frac{dI}{dE})$
,$I_{1}$$(\equiv \frac{2\pi\hbar}{e}\frac{dI_{1}}{dE})$
and $I_{2}$$(\equiv \frac{2\pi\hbar}{e}\frac{dI_{2}}{dE})$
and we have set units of $\hbar$ and 2m to be unity.

III {\bf Results and discussions}

In the limit V=0 our expressions agree with earlier known
results[19,20]. We have studied the behavior of $I_1$ and $I_2$
as a function of Fermi wavevectors and identify the wave vector
intervals, wherein either the $I_1$ or $I_2$ flows in the
negative direction for $V \ne 0$.  In any range of Fermi energy
if one of them ($I_1$ or $I_2$) is negative then the magnitude
of the negative current gives the circulating current. When
$I_1$ is negative the direction of circulating current is
anticlockwise and when $I_2$ is negative then it is clockwise.
Clockwise circulating current is taken to be positive and
anticlockwise as negative according to the usual convention
followed for persistent currents in closed rings.  Having
defined circulating current in this manner we plot in fig.  (2)
circulating current versus kL for two values of the impurity
potential strength. The solid curve is for VL=0 and the dotted
curve is for VL=1. In both the cases we take $l_1/L$=.3125,
$l_2/L$=.3125 and $l_3/L$=.375. The figure shows that the dotted
curve is slightly shifted towards higher energy axis with
respect to the solid curve and the first peak value of it is
larger than that of the solid curve. It shows that in the first
energy range where we obtain circulating current the amplitude
of the current is actually enhanced by the impurity potential
and the position of this range of Fermi energy is modified due
to the impurity. Amplitude of the first negative peak in the
circulating current decreases, and the circulating current
behavior in the interval of kL between 10-15 is qualitatively
modified. As discussed in our earlier paper[19] the nature of
the circulating current depends on the zero-pole pair structure
in the transmission amplitude. The circulating current arises
near the poles (or at a real value of the pole in the complex
plane) in the transmission amplitude.  Imaginary value of the
poles in the complex plane determines the width of the peak. The
poles determine the resonant states of the system. The smaller
is the imaginary value the narrower is the peak.  The shift in
the peak position on the real axis is attributed to the fact
that the position of the poles change as we change the impurity
potential.  circulating current is also enhanced at some other
Fermi energy ranges whereas at some other it is suppressed
because of the impurity. This happens because of the fact that
circulating current due to current magnification has a
completely different origin than that of persistent currents due
to the presence of magnetic field in closed isolated rings.

Several circulating current peaks exhibit exotic behavior as we
keep increasing the impurity strength. For the sake of clarity
we will consider only the behavior of circulating currents in
the first range of kL around which circulating current arises.
In fig (3) we plot circulating current versus kL for various
values of VL in the first energy range which keeps changing with
change in VL. The curves a, b, c, d and e are for VL=5, 10, 15,
20 and 25, respectively. For all the curves we take
$l_1/L$=.3125, $l_2/L$=.3125 and $l_3/L$=.375.  The figure shows
that initially the magnitude of the first peak on the positive
side increases monotonously and also the peak position shifts
towards higher energy with VL.  However, as the magnitude of the
peak value at the maximum increases the width of the peak
decreases.  On the other hand the magnitude of the circulating
current peak on the negative side in this first energy range is
not so sensitive to the impurity strength VL.  Initially as VL
increases it decreases slightly. This can be seen by observing
the curves a, b, c and d. For curve e this peak on the negative
side has increased slightly with respect to the curve d. So this
negative side peak exhibits a small oscillatory behavior with
VL. However, the width of this negative peak monotonously
decreases as a function of the strength of the impurity in the
given range of potential strength considered in the figure.
Needless to say that the shape of the peaks also reflects the
nature of density of states around the position of the peak (or
at resonance). Enhancement of density of states around the poles
leads to large circulating currents.  This can be easily
verified by calculating the density of states near the poles by
using Friedel's theorem[26], according to which change in
density of states due to a scatterer, at a particular energy is
given by ${1 \over \pi}{d \theta \over dE}$, where $ \theta$ is
the arguement of the complex transmission amplitude.

We from now on study the behavior of the maximum value
($I_{max}$) of the first circulating current peak on the
positive side as a function of the impurity strength. The
position of the $I_{max}$ on the real kL axis is given by the
real part of the pole in the transmission amplitude (or
scattering amplitude) in a complex kL plane. It is clear from
the earlier fig. 3 that $I_{max}$ increases initially as a
function of VL. In fig. 4 we have plotted $I_{max}$ versus VL,
the values of the physical parameters $l_1/L$, $l_2/L$ and
$l_3/L$ are same as that used for fig. 3.  We numerically
evaluate $I_{max}$ of circulating current (at the first peak)
for a fixed value of VL and then slowly change VL upto VL=31 in
order to plot the peak value at the maximum of circulating
current versus VL. In fig. 4 we have not shown the behavior of
$I_{max}$ in the range VL=(28.0) to VL=(29.1) as the magnitude
of the $I_{max}$ increases rapidly and goes beyond the scale.
Between a value of VL=28 and VL=29 the $I_{max}$ is several
orders of magnitude larger than the total current I through the
sample and the corresponding widths of the peaks are extremely
small. For example $I_{max}$=111.123, 175.687, 709.423,
8958.1428 and 11755 for values of VL=28.2, 28.5, 28.9, 28.99 and
29.01, respectively and the peak value $I_{max}$ on kL axis
corresponds to kL=8.3549242, 8.3633125, 8.374304, 8.37674705 and
8.377288471, respectively.  In fact $I_{max}$ diverges for
VL=(29) (for this case corresponding kL=8.37701785204). Above
VL=29 the $I_{max}$ shows a drastic fall in the magnitude and as
expected it reaches the value zero in the limit VL$\rightarrow
\infty$.  In this limit the current in the upper arm is zero
($I_1=0$) and $I_2$=I and hence no circulating current flows.
Our above observations clearly indicate that there is no upper
bound for the current magnification (same as in the case for
classical LCR networks driven by external a.c. electromotive
force[22] in the limit the magnitude of resistance R
$\rightarrow$ 0). For the particular case considered here as we
increase impurity strength initially it helps the current
magnification and on reaching a critical value (at which current
magnification diverges) further increase in impurity strength
decrease the current magnification. Thus impurity in general can
play a dual role of enhancing as well as suppressing the current
magnification effect.

To analyze in some details the divergence in current
magnification, we study the imaginary part of the pole $I_m(p)$
in the complex kL plane.  In fig. (5) we have plotted the
imaginary part of the pole (first pole) corresponding to fig.
(4) (by analyzing the transmission amplitude in the complex kL
plane) as a function of VL.  All the physical parameters are the
same as that used for fig.  (4). One can immediately notice that
the imaginary part of the pole decreases as we increase VL. This
corresponds to a situation where the current magnification
increases as a function of VL. At a value of VL =(29) imaginary
part approaches zero. This is the same value of VL where the
current magnification exhibits divergence. In a sense the
divergence of current magnification occurs as poles of a
transmission amplitude in the complex plane approaches the real
axis. The lifetime of the resonant state (corresponding to
poles) is inversely proportional to the imaginary part of the
pole. The imaginary part being zero implies that the resonant
state has an infinite lifetime, and hence is a bound state. This
bound state is localized across the loop and corresponds to a
von Neumann and Wigner type bound state in continuum[24,27].  As
we increase VL further the imaginary part increases and
consequently the peak value of circulating current decreases.

The circulating current due to current magnification generally
appears around the zeros of the transmission coefficient but
there can be  cases, where this is not true as shown in fig.
(6). Here we have plotted circulating current (solid curve) and
the transmission coefficient (dashed curve) versus kL for VL=20,
$l_1/L$=.3125, $l_2/L$=.3125 and $l_3/L$=.375. In the kL range
from 34 to 45 we find a large circulating current and  around
this region T goes to zero nowhere but has a minimum
(antiresonance).  As mentioned in the introduction in the
special case of a metal loop with two equal arms and without the
presence of impurity, the currents $I_1$ and $I_2$ on the arms
(due to symmetry reasons) are equal and each equal half of I at
all Fermi-energies.  So at no Fermi energy interval we get a
circulating current according to our definition for it as stated
earlier.  However, a small impurity in one of the arms breaks
the above symmetry and the currents $I_1$ and $I_2$ are not
equal. This gives rise to circulating current in certain range
of the Fermi energies, and we find that as the impurity strength
is increased from zero, initially the circulating current
amplitudes in all the relevant kL intervals show tendency to
increase.

So far we have discussed the effect of the impurity strength on
the current magnification. In the following we consider the
effect of the impurity position on the current magnification.
For this we consider a fixed impurity strength and fixed upper
and lower arm lengths. In fig. (7) we have plotted the
circulating currents as a function of kL for a fixed value of
impurity strength VL=(10) and the impurity placed at various
different positions in the upper arm.  The solid curve is the
case when the impurity is placed in the middle of the upper arm
($l_1/L$=.3125, $l_2/L$ =.3125 and $l_3/L$=.375) whereas the
dashed curve is for the impurity position away from the center
of the arm ($l_1/L$=.4, $l_2/L$ =.225 and $l_3/L$=.375).  We
observe drastic changes in the nature, height and width of the
circulating current peaks in the two cases. Again, this can be
explained from the point of view of the shift in the pole
structure for the two different cases.

In conclusion, we have studied the effect of impurity on the
current magnification in an open metallic loop in the presence
of a transport current. In this non-equilibrium steady state
circulating current arises even in the absence of magnetic
field. More importantly we have shown that the presence of
impurity can  dramatically enhance the current magnification
effect at particular values of Fermi energies, whereas it can
decrease current magnification at some other values of
Fermi-energies. We have shown that current magnification
property is not only sensitive to the impurity strength but also
sensitive to its position. This is in contrast to the effect of
the presence impurity on persistent currents in closed isolated
metallic rings, where persistent currents are always suppressed
as one increases the impurity strength.  Since the underlying
principle of circulating currents in open systems in the
presence of a transport current (and in the absence of magnetic
field) is different from the persistent currents in isolated
rings in the presence of magnetic flux, we expect that even in
the presence of random impurities one should expect current
magnification of observable magnitudes in open systems in the
presence of a transport current. Work along these lines is in
progress. The circulating current changes sign if we change the
direction of the current and hence at equilibrium ($\mu_1 =
\mu_2$) no circulating currents are possible.
Since the magnetic moment of the loop is proportional to the
line integral of the current (i.e., the total current integrated
over Fermi energies between $\mu_1$ and $\mu_2$ at temperature
T=0) along the entire circumference of the loop, due to the
current magnification effect we expect that one should observe
enhanced magnetic response around particular Fermi energy
intervals. This can be achieved experimentally by having a gate
(which mimicks the impurity potential) in one of the arms and by
appropriately tuning the gate voltage and Fermi energy or
($\mu_1 - \mu_2$).  The experiment will show dramatic
enhancement in the magnetic response. The effect of the magnetic
field on the current magnification will be presented in our
forthcoming publication.

{\bf Acknowledgements}

One of us (AMJ) thanks Professors A. G. Aranov and V. E.
Kravtsov for several useful discussions during the initial
stages of this work and Professor N. Kumar for continued
discussions on this subject.

\vfill
\eject

\vfill
\eject
{\bf Figure captions}

Fig. 1. Metal loop connected to two electron
reservoirs with chemical potentials $\mu_1$ and $\mu_2$. There
is a delta function potential impurity at the site X in the upper arm.
Fig. 2. Plot of circulating current versus kL for VL=0 (solid
line) and VL=1 (dotted line). In both the cases $l_1/L$=.3125,
$l_2/L$=.3125 and $l_3/L$=.375.

Fig. 3. Plot of circulating current versus kL for various values
of VL in the first energy range. The curves a, b, c, d and e are
for VL=5, 10, 15, 20 and 25, respectively. For all the curves $l_1/L$=.3125,
$l_2/L$=.3125 and $l_3/L$=.375.

Fig. 4. The figure shows the scaling of $I_{max}$ with VL for $l_1/L$=.3125,
$l_2/L$=.3125 and $l_3/L$=.375.

Fig. 5. The plot of strength or magnitude of the imaginary part
of the first complex pole of the transmission amplitude versus
VL for $l_1/L$=.3125,
$l_2/L$=.3125 and $l_3/L$=.375.

Fig. 6. Plot of circulating current (solid curve) and
transmission coefficient (dashed curve) versus kL for VL=20, $l_1/L$=.3125,
$l_2/L$=.3125 and $l_3/L$=.375.

Fig. 7. Plot of circulating current versus kL for VL=(10) and for
two positions, respectively,
of the impurity in the upper arm. The solid curve
is for $l_1/L$=.3125,
$l_2/L$=.3125 and $l_3/L$=.375. The dotted curve is for $l_1/L$=.4,
$l_2/L$=.225 and $l_3/L$=.375.

\vfill
\eject
\end{document}